\documentstyle[12pt]{article}
\begin{document}
\begin{flushright}
hep-th/0204015\\
SNBNCBS-2002
\end{flushright}
\vskip 3cm
\begin{center}
{\bf \Large { Superfield approach to topological 
features of \\non-Abelian gauge theory}}

\vskip 3cm

{\bf R.P.Malik}
\footnote{ E-mail address: malik@boson.bose.res.in  }\\
{\it S. N. Bose National Centre for Basic Sciences,} \\
{\it Block-JD, Sector-III, Salt Lake, Calcutta-700 098, India} \\

\vskip 3.5cm

\end{center}

\noindent
{\bf Abstract}:
We discuss some of the key topological aspects of a two $(1+1)$-dimensional 
(2D) self-interacting non-Abelian gauge theory (having no interaction with 
matter fields) in the framework of {\it chiral} superfield formalism. We 
provide the geometrical interpretation for the Lagrangian density, symmetric 
energy momentum tensor, topological invariants, etc., by exploiting the 
{\it on-shell} nilpotent BRST- and co-BRST symmetries that emerge after the 
application of (dual) horizontality conditions. We show that the above 
physically interesting quantities geometrically correspond to the translation 
of some local (but composite) {\it chiral} superfields along one of the two 
independent Grassmannian directions of a four ($2+2)$-dimensional 
supermanifold. This translation is generated by the conserved and on-shell 
nilpotent (co-)BRST charges that are present in the theory. 

\vskip 1.0cm


\baselineskip=16pt

\vskip 1cm

\newpage

\noindent
{\bf 1 Introduction}\\

\noindent
The modern developments in the subject of
topological field theories (TFTs) have encompassed in their ever-widening 
horizons a host of diverse and distinct areas of research in theoretical 
physics and mathematics. In this context, mention can be made of such 
interesting topics as Chern-Simon theories, topological string 
theories and matrix models, 2D topological gravity, Morse theory, Donaldson 
and Jones polynomials, etc. (see, e.g., Ref. [1] and references therein for 
details). Without going into the subtleties and intricacies, TFTs can be 
broadly classified into two types. The Witten type TFTs [2]
are the ones where the Lagrangian density turns out to be the 
Becchi-Rouet-Stora-Tyutin (BRST) (anti-)commutator. The conserved and
nilpotent BRST charge for such a class of
TFTs generates a symmetry that is a combination
of a topological shift symmetry and some types of local gauge symmetry. On the
other hand, the Schwarz type of TFTs [3] are characterized by the existence
of a conserved and nilpotent BRST charge that generates only some local gauge
type of symmetries for a Lagrangian density that {\it cannot} be 
totally expressed as
the BRST (anti-)commutator (see, e.g., Ref. [1] for details). For both types
of TFT, there are no energy excitations in the physical sector of the theory
because the energy-momentum tensor turns out to be a BRST
(anti-)commutator and all the physical states (including the vacuum) of this 
theory are supposed to be invariant w.r.t. the conserved, nilpotent,
metric independent and hermitian BRST charge $Q_{b}$
(i.e. $Q_{b} |vac> = 0, Q_{b} |phys> = 0$). 

Recently, in a set of papers [4-7], the free 2D Abelian- and self-interacting 
non-Abelian gauge theories (without any kind of interaction with matter fields) have been shown to belong to a new class of TFTs
because the Lagrangian density of the theory turns out to bear an appearance
similar to the Witten type theories but the local symmetries of the theory are
that of Schwarz type. Furthermore, these non-interacting 
as well as self-interacting 2D theories
[4-7], interacting 2D Abelian gauge theory (where there is an 
interaction between Dirac fields and 2D photon) [8,9] and 
$(3 + 1)$-dimensional (4D) free Abelian
two-form gauge theory [10] have been shown to provide a set of tractable
field theoretical models for the Hodge theory where the local, covariant
and continuous symmetries of the Lagrangian density and corresponding 
conserved charges (as generators) are identified with all
the three de Rham cohomology operators of differential geometry
\footnote{ On a compact manifold without a boundary, the set of operators 
$(d, \delta, \Delta)$ (with $ d = dx^\mu \partial_\mu, \delta = \pm * d *, 
\Delta = d \delta + \delta d)$ define the de Rham cohomological properties 
of the differential forms. They are called the exterior 
derivative, co-exterior derivative and Laplacian operator respectively
and obey: $ d^2 = \delta^2 = 0, [ \Delta, d ] = [ \Delta, \delta ]
= 0, \Delta = \{ d, \delta \} = (d + \delta)^2$. Here $*$ stands for
the Hodge duality operation [11-14].}. The 
geometrical interpretation for these charges as the translation generators 
along some specific directions of a four
$(2 + 2)$-dimensional supermanifold has also been established 
for the 2D free- and self-interacting (non-)Abelian gauge theories [15-17].
In a recent paper [18], some of the key features
of the topological nature of a 2D free Abelian gauge theory have been captured
in the superfield formulation [19-23] and their geometrical interpretations 
have been provided in the language of translations along some specific
directions of the $(2 + 2)$-dimensional supermanifold.
One of the central themes of our present paper is to extend 
our work on the free 2D Abelian gauge theory [18] to the 
{\it more general case } of self-interacting
2D non-Abelian gauge theory and provide the geometrical interpretation for
some key topological properties associated with this theory in the
framework of (geometrical) superfield formulation [19-23]. Such studies
are important because they provide geometrical origin for some of the key
topological quantities of physical interest (e.g. Lagrangian density,
symmetric energy momentum tensor, topological invariants, etc.) for the 
(non-)Abelian gauge theories. In particular, our results on the geometrical
origin and interpretation for the Lagrangian density and corresponding 
symmetric energy momentum tensor are completely novel in nature 
{\it vis-{\`a}-vis} the key results of Refs. [20-22,17] where the
geometrical interpretation for {\it only} the nilpotent (anti-)BRST charges 
[20-22,17] (and (anti-)co-BRST charges [17]) has been provided. The key
observations of our present paper are, however, similar to [18].

The self-interacting non-Abelian gauge theory (having no interaction with 
matter fields) is described by 
a singular Lagrangian density that happens to be endowed with
first-class constraints in the language of Dirac's classification scheme
[24,25]. For the BRST quantization of such a theory, the original Lagrangian
density is extended to include the gauge-fixing- and Faddeev-Popov ghost terms
so that the theory can maintain  unitarity and
``quantum'' gauge symmetry (i.e. nilpotent BRST symmetry) {\it together}
 at any arbitrary order of perturbation theory. The ensuing Lagrangian
density that respects the {\it on-shell} nilpotent BRST symmetry, however,
does not respect corresponding on-shell nilpotent anti-BRST symmetry. 
To the best of our familiarity with the relevant literature,
the on-shell nilpotent version of the anti-BRST symmetry does not 
exist for the self-interacting and/or interacting non-Abelian gauge theory 
in any arbitrary spacetime dimension. This
feature of the self-interacting non-Abelian gauge theory is drastically
different from its Abelian counterpart where both the on-shell nilpotent
(anti-)BRST symmetries are respected by the one and the same Lagrangian 
density (see, e.g.,[15,16]). 
A possible explanation for this discrepancy has been provided
in a recent paper [26] by resorting to the geometrical superfield approach 
to BRST formalism. In our present paper, we obtain the
on-shell nilpotent version of the BRST- and co-BRST symmetries 
for the 2D self-interacting non-Abelian gauge theory by exploiting the
generalized versions of horizontality condition
\footnote{ This condition is referred to as the ``soul-flatness'' condition
by Nakanishi and Ojima [27].}
w.r.t. the (super) cohomological operators $(\tilde d) d$ (together with the 
Maurer-Cartan equation) and $(\tilde \delta) \delta$
defined on the $(2 + 2)$-dimensional supermanifold. In this endeavour, the
choice of the superfields to be chiral plays a very decisive role as the
(dual) horizontality conditions (w.r.t. $(\tilde \delta)\delta$
and $(\tilde d) d$) lead to the derivation of the on-shell nilpotent (co-)BRST
symmetries {\it only}. The off-shell version
of these symmetries has been already obtained in Ref. [17] where the ideas of
Refs. [20-22] on the (anti-)BRST symmetries have been expanded and
new (anti-)co-BRST symmetries have been 
introduced and derived in the superfield formalism by exploiting the above
(super) cohomological operators together with the imposition of (dual)
horizontality conditions. In contrast to the choice of chiral superfields
for the derivation of the on-shell nilpotent (co-)BRST symmetries,
the off-shell nilpotent (anti-)BRST- and (anti-) co-BRST
symmetries have been derived by taking into account
the most general superfield expansion along the
$\theta$, $\bar \theta$ and $\theta\bar\theta$-directions of the
$(2 + 2)$-dimensional supermanifold [17].

In our present discussion, we concentrate only
on the on-shell version of nilpotent BRST- and co-BRST symmetries
(avoiding any discussion about anti-BRST- and anti-co-BRST symmetries) because
the basic Lagrangian density (see, e.g., (2.1) below) respects only these
symmetries. The derivation of these symmetries and corresponding generators
are good enough to shed some light on the topological nature of the 2D
self-interacting non-Abelian gauge theory. In fact,
the topological nature of this theory is 
encoded in the form of the Lagrangian density and the symmetric 
energy-momentum tensor which can be thought of as the translation of some
local (but composite) chiral superfields along one of the two Grassmannian 
directions of the supermanifold. This translation is 
generated by the on-shell nilpotent BRST- and co-BRST charges which turn out 
to geometrically correspond to the translation generators along
one of the two Grassmannian directions of the
four $(2+2)$-dimensional supermanifold. In mathematical terms, the Lagrangian
density and symmetric energy momentum tensor for the present theory turn
out to be the total derivative of some local (but composite) chiral superfields
w.r.t. one of the two Grassmannian variables (cf. (5.1) and (5.5) below). In
these derivations, the chiral superfield expansions are taken to be the ones
that are obtained after the application of (dual) horizontality conditions.
Thus, the (super) cohomological operators $(\tilde \delta)\delta$ and
$(\tilde d) d$ play very important and pivotal roles for our
present discussions through the (dual) horizontality restrictions.

Our present investigation is essential primarily on three counts. First of
all, to the best of our knowledge, the full potential of 
(super) co-exterior derivatives $(\tilde \delta) \delta$
has not yet been thoroughly exploited in the
context of superfield approach to BRST formalism except in some of our
recent works [15-18]. Thus, besides whatever have been achieved and understood
in [15-18,26], it is important to explore the utility of these
(super) cohomological operators in their diverse, distinct and multiple forms. 
Second, our present paper
explains the reason behind the existence of on-shell nilpotent (co-)BRST 
symmetries for the Lagrangian density (cf. (2.1) below) that does
not respect on-shell nilpotent anti-BRST and anti-co-BRST symmetries. In fact, 
the choice of the chiral superfields (along with the idea of (dual) 
horizontality conditions) plays an important role in proving  the existence 
of on-shell nilpotent (co-)BRST symmetries. The choice of the anti-chiral 
superfields does not lead to the derivation of anti-BRST and anti-co-BRST 
symmetries for this theory as explained in our recent work [26]. Finally, 
the  geometrical understanding of the Lagrangian density and the symmetric 
energy momentum tensor
for the present theory {\it might} turn out to be useful in the understanding 
of topological 2D gravity and topological string theories
where a non-trivial
metric is chosen for the theoretical discussions of such kind of gauge theories 
in the background of curved spacetime.

The contents of our present paper are organized as follows. In section 2, we
set up the notations and briefly recapitulate the bare essentials of the 
BRST- and co-BRST symmetries for the 2D
self-interacting non-Abelian gauge theory in the Lagrangian formulation.
Sections 3 and 4 are devoted to the derivation of the above on-shell nilpotent
symmetries in the framework of superfield formalism. In section 5, we discuss
topological aspects and provide their geometrical interpretations in the 
language of translations on the $(2+2)$-dimensional supermanifold.
Finally, in section 6, we make some concluding remarks and point out a few
directions that can be pursued later.\\

\noindent
{\bf 2 BRST- and co-BRST symmetries: Lagrangian formulation}\\

\noindent
Let us begin with the BRST invariant Lagrangian density ${\cal L}_{B}$
for the self-interacting  two ($1 + 1)$-dimensional
\footnote{We adopt here the conventions and notations such that the 2D flat
Minkowski metric is: $\eta_{\mu\nu} =$ diag $(+1, -1)$ and $\Box = 
\eta^{\mu\nu} \partial_{\mu} \partial_{\nu} = (\partial_{0})^2 - 
(\partial_{1})^2,  F_{01} = F^{10}
= -\varepsilon^{\mu\nu} (\partial_\mu A_\nu + \frac{1}{2}\; A_\mu \times A_\nu)
= E = \partial_{0} A_{1} - \partial_{1} A_{0} + A_{0} \times A_{1}, 
\varepsilon_{01} = \varepsilon^{10} = + 1, \; 
D_\mu C = \partial_\mu C + A_\mu \times C, 
\alpha \cdot \beta = \alpha^a \beta^a , 
(\alpha \times \beta)^a = f^{abc} \alpha^b \beta^c$ for non-null vectors 
$\alpha$ and $\beta$ in the group space. Here Greek 
indices: $\mu, \nu...= 0, 1$ correspond to the spacetime directions on
the 2D manifold and Latin indices: $ a, b, c..= 1, 2, 3...$ stand
for the Lie group ``colour'' values.}
non-Abelian gauge theory in the Feynman gauge [27-30]
$$
\begin{array}{lcl}
{\cal L}_{B} &=& - \frac{1}{4}\; F^{\mu\nu} \cdot F_{\mu\nu} 
- \frac{1}{2}\; (\partial_\mu  A^\mu)\cdot (\partial_\rho A^\rho)
- i \;\partial_{\mu} \bar C  \cdot D^\mu C \nonumber\\
 &\equiv& \frac{1}{2}\; E \cdot E
- \frac{1}{2} \; (\partial_\mu  A^\mu) \cdot (\partial_\rho A^\rho)
- i \;\partial_{\mu} \bar C \cdot D^\mu C
\end{array} \eqno(2.1)
$$
where $F_{\mu\nu} = \partial_\mu A_\nu - \partial_\nu A_\mu
+ A_\mu \times A_\nu$ is the field
strength tensor derived from the connection one-form $ A = dx^\mu A_\mu^a T^a$
(with $A_\mu = A_\mu^a T^a$ as the vector potential) by application of the 
Maurer-Cartan equation $ F = d A + A \wedge A$ where $d = dx^\mu 
\partial_\mu $ is the exterior derivative (with $d^2 = 0)$ and 
the two-form $ F = \frac{1}{2}\; (dx^\mu \wedge dx^\nu) F_{\mu\nu}^a T^a$.
In 2D spacetime, only the electric field component
($F_{01} = E = E^a T^a $) of the field strength tensor 
$F_{\mu\nu}$ exists.
Here $T^a$ form the compact Lie algebra: $[ T^a , T^b ] = f^{abc} T^c$
with structure constants $f^{abc}$ which can be chosen to be totally
antisymmetric in $a,b,c$ (see, e.g., Ref. [30] for details). The 
gauge-fixing term is derived as $\delta A = (\partial_\rho A^{\rho a} T^a)$ 
where $\delta = - * d * $ (with
$\delta^2 = 0$) is the co-exterior derivative and $*$ is the Hodge duality
operation. The (anti-)commuting ($C^a \bar C^b + \bar C^b C^a = 0, 
(C^a)^2 = (\bar C^a)^2 = 0)$ (anti-)ghost fields $(\bar C^a)C^a$ are required 
in the BRST invariant theory to maintain unitarity and ``quantum'' gauge 
(i.e. BRST) invariance together at any arbitrary order of perturbative
calculations. In fact, these (anti-)ghost fields (which are not matter fields)
interact with the self-interacting non-Abelian gauge fields 
($A_\mu = A_\mu^a T^a$) only in the loop diagrams of perturbation theory
(see, e.g., Ref. [31] for details). The above Lagrangian density
(2.1) respects ($s_{b} {\cal L}_{B} = - \partial_\mu [ (\partial_\rho A^\rho)
\cdot D^\mu C ], s_{d} {\cal L}_{B} = \partial_\mu 
[ E \cdot \partial^\mu \bar C ]$)
the following on-shell ($ \partial_\mu D^\mu C 
= D_\mu \partial^\mu \bar C = 0$)
nilpotent $(s_{b}^2 = 0,  s_{d}^2 = 0)$ BRST ($s_{b}$)
\footnote{We follow here the notations and conventions of Ref. [30]. In fact,
in its full glory, a nilpotent ($\delta_{(D)B}^2 = 0$)
(co-)BRST transformation $(\delta_{(D)B})$ is equivalent 
to the product of an 
anti-commuting ($\eta C = - C \eta, \eta \bar C = - \bar C \eta$)
spacetime independent parameter $\eta$ and $(s_{d})s_{b}$ 
(i.e. $\delta_{(D)B} = \eta \; s_{(d)b}$) where $s_{(d)b}^2 = 0$.} 
-and dual(co)-BRST ($s_{d}$) symmetry transformations [5,6,17,26]:
$$
\begin{array}{lcl}
s_{b} A_{\mu} &=& D_{\mu} C \quad s_{b} C = -\frac{1}{2} C \times C
\quad s_{b} \bar C = -i (\partial_\mu A^\mu) \quad
s_{b} E = E \times C  \nonumber\\
s_{d} A_{\mu} &=& - \varepsilon_{\mu\nu} \partial^\nu \bar C \qquad
s_{d} \bar C = 0 \qquad  s_{d} C = - i E \qquad s_{d} E = D_\mu 
\partial^\mu \bar C.
\end{array}\eqno(2.2)
$$
The above continuous symmetries, according to Noether's theorem, lead to
the following expressions for the conserved and on-shell nilpotent 
(co-)BRST charges $(Q_{d})Q_{b}$ [5,6]
$$
\begin{array}{lcl}
Q_{b} &=& {\displaystyle \int}\; dx\;
\bigl [\;\partial_{0} (\partial_\rho A^\rho) \cdot C -
(\partial_\rho A^\rho) \cdot D_{0} C + \frac{i}{2}\; \dot {\bar C} \cdot
C \times C \;\bigr ] \nonumber\\
Q_{d} &=& {\displaystyle \int}\; dx\;
\bigl [\;E \cdot \dot {\bar C}  -
D_{0} E \cdot {\bar C} -  i\; \bar C  \cdot \partial_{1} \bar C \times C
\;\bigr ]
\end{array} \eqno(2.3)
$$
which turn out to be the generator for the transformations (2.2). This
latter statement can be succinctly expressed 
in the mathematical form (for the generic field $\Psi = \Psi^a T^a$) as
$$
\begin{array}{lcl}
s_{r} \Psi = - i \; [\; \Psi, Q_{r} \;]_{\pm} \;\;\; \qquad \;\;\;
r = b,  d 
\end{array} \eqno(2.4)
$$
where brackets $[\;, \;]_{\pm}$ stand for the (anti-)commutators for 
any arbitrary generic field $\Psi (\equiv A_\mu, C , \bar C)$ 
being (fermionic)bosonic in nature. Left to itself, the Lagrangian density
(2.1) does {\it not} respect any anti-BRST- and anti-co-BRST symmetries.
These symmetries can be brought in, however, by modifying (2.1) to incorporate
a specific set of auxiliary fields (see, e.g., Refs. [5,6]).
It is interesting to note that only the {\it off-shell nilpotent} version of 
these symmetries exist for the modified Lagrangian density
 (see, e.g., Ref. [27-30] for details). 
With the help of equations (2.3) and (2.4), the Lagrangian density (2.1)
can be expressed, modulo some total derivatives, as the sum of 
on-shell nilpotent BRST- and co-BRST anti-commutators:
$$
\begin{array}{lcl}
{\cal L}_{B} &=& \{ Q_{d} , \;\frac{1}{2} E \cdot C \}
- \{ Q_{b} , \;\frac{1}{2}\;(\partial_\rho A^\rho) \cdot \bar C \}
\equiv s_{d}\;  [\; \frac{i}{2} E \cdot C\;  ] - s_{b}
\; [\; \frac{i}{2} (\partial_\rho A^\rho) \cdot \bar C \; ].
\end{array} \eqno(2.5)
$$
The appearance of the above Lagrangian density is that of Witten type TFTs
when the physical states (and vacuum) of the theory are 
supposed to be annihilated by 
$Q_{b}$ and $Q_{d}$. Such situation does arise if we invoke harmonic
state of the Hodge decomposed state to correspond to the 
{\it physical state}  in the
total quantum Hilbert space [5-7].
The expression for the symmetric energy-momentum tensor $T_{\mu\nu}^{(s)}$
for the Lagrangian density (${\cal L}_{B}$) is
$$
\begin{array}{lcl}
T_{\mu\nu}^{(s)} &=& 
- \frac{1}{2} (\partial_\rho A^\rho) \cdot
(\partial_\mu A_\nu + \partial_\nu A_\mu)
- \frac{1}{2} E \cdot
(\varepsilon_{\nu\rho} \partial_\mu A^\rho 
+ \varepsilon_{\mu\rho} \partial_\nu A^\rho) \nonumber\\
&-& \frac{i}{2} (\partial_\mu \bar C) \cdot
(D_\nu C  + \partial_\nu C)
- \frac{i}{2} (\partial_\nu \bar C) \cdot
(D_\mu C  + \partial_\mu C) - \eta_{\mu\nu} {\cal L}_{B}.
\end{array} \eqno(2.6)
$$
This also turns out, modulo some total derivatives,  to be the sum of
BRST and co-BRST anti-commutators as given below [6]
$$
\begin{array}{lcl}
T_{\mu\nu}^{(s)} &=& 
\{ Q_{b} , L_{\mu\nu}^{(1)} \} +
\{ Q_{d} , L_{\mu\nu}^{(2)} \} \equiv s_{b} \;(i L_{\mu\nu}^{(1)})
+ s_{d}\; (i L_{\mu\nu}^{(2)}) \nonumber\\
L_{\mu\nu}^{(1)} &=& \frac{1}{2}\; \bigl [\; \partial_\mu \bar C \cdot A_\nu
+ \partial_\nu \bar C \cdot A_\mu + \eta_{\mu\nu} (\partial_\rho A^\rho)
\cdot \bar C \;\bigr ] \nonumber\\
L_{\mu\nu}^{(2)} &=& \frac{1}{2}\; \bigl [\; \partial_\mu C \cdot 
\varepsilon_{\nu\rho} A^\rho +
\partial_\nu C \cdot \varepsilon_{\mu\rho} A^\rho - \eta_{\mu\nu} 
E \cdot C \;\bigr ].
\end{array} \eqno(2.7)
$$
The generic form of the topological invariants ($I_{k}$ and
$J_{k}$) with respect to the conserved and on-shell
($\partial_\mu D^\mu C = D_\mu \partial^\mu \bar C = 0$) nilpotent
($Q_{b}^2 = Q_{d}^2 = 0$) BRST ($Q_{b}$) - and co-BRST  ($Q_{d}$) charges,
on the 2D manifold, are
$$
\begin{array}{lcl} 
I_{k} = {\displaystyle \oint}_{C_{k}} \; V_{k} \qquad 
J_{k} = {\displaystyle \oint}_{C_{k}} W_{k}
\qquad (k = 0, 1, 2)
\end{array} \eqno(2.8)
$$
where $C_{k}$ are the $k$-dimensional homology cycles in the 2D manifold 
and $V_{k}$ and $W_{k}$ are the $k$-forms w.r.t. $Q_{b}$ and $Q_{d}$
respectively
\footnote{It will be noted that we have chosen a 2D Minkowskian manifold
where metric has opposite signs in the diagonal entries. To be very precise,
this manifold is not a compact manifold. To have the accurate meaning
of the topological invariants, homology cycles, etc., (and their connections 
with the notions in the algebraic geometry), one has to consider the Euclidean 
version of the 2D Minkowskian manifold  which turns out to be
the 2D closed Riemann surface.
In this case, the 2D metric will have the same signs in the diagonal
entries and $\mu, \nu, \rho...= 1, 2$. In fact,  on these lines,
a detailed analysis for the 2D (non-)Abelian
gauge theories has been performed in Ref. [32]. 
For the sake of brevity, however, we shall
continue with our Minkowskian notations but we shall keep in mind
this crucial point and decisive argument.}. These forms are [5,6]
$$
\begin{array}{lcl} 
V_{0} &=& - (\partial_\rho A^\rho) \cdot C - \frac{i}{2}\; \bar C \cdot 
C \times C \qquad V_{1} = [\; - (\partial_\rho A^\rho) \cdot A_\mu 
+ i\; C \cdot D_\mu \bar C \;]\; dx^\mu \nonumber\\
V_{2} &=& i \; [\;A_\mu \cdot D_\nu \bar C - \bar C \cdot D_\mu A_\nu \;]
\;dx^\mu \wedge dx^\nu
\end{array} \eqno(2.9)
$$
$$
\begin{array}{lcl} 
W_{0} &=& E \cdot \bar C \qquad W_{1} = [\; \bar C \cdot \varepsilon_{\mu\rho}
\partial^\rho C - i\; E \cdot A_\mu\; ]\; dx^\mu \nonumber\\
W_{2} &=& i\; [\; \varepsilon_{\mu\rho} \partial^\rho C \cdot A_\nu
+ \frac{1}{2}\; C \cdot \varepsilon_{\mu\nu} (\partial_\rho A^\rho) \;]\;
dx^\mu \wedge dx^\nu.
\end{array} \eqno(2.10)
$$
These topological invariants obey certain specific kind of
recursion relations [5,6]
which primarily shed some light on the connection between 
(co-)BRST transformations- {\it and} operation of (co-)exterior derivatives
on these invariants (cf. (5.4) below). Equations (2.5)-(2.10) establish the
topological nature of the above self-interacting 2D non-Abelian gauge theory.
For the case of this theory with off-shell nilpotent (anti-)BRST and
(anti-)co-BRST symmetries and corresponding conserved and 
off-shell nilpotent charges, a set of four 
topological invariants has been computed in [6]. In what follows hereafter, 
we shall concentrate, however, only on the Lagrangian density (2.1), its 
on-shell nilpotent symmetry generators (i.e. (co-)BRST charges)
 and corresponding topological invariants.\\

\noindent
{\bf 3 On-shell nilpotent BRST symmtery: superfield formulation}\\

\noindent
To provide the geometrical interpretation for the conserved and on-shell
nilpotent BRST charge $Q_{b}$ (cf. (2.3)) as the translation generator in the
framework of the superfield formulation [19-23], we first generalize the
basic generic local field $\Psi (x) = (A_\mu (x), C (x), \bar C (x))$ of 
the Lagrangian
density (2.1) to a chiral ($\partial_{\theta} V_{s} (x,\theta,\bar\theta)
= 0$) supervector superfield $V_{s} = (B_\mu (x, \bar \theta), \Phi
(x, \bar\theta), \bar \Phi (x,\bar\theta))$ defined on the 
$(2+2)$-dimensional supermanifold with the following super expansions
along the Grassmannian direction $\bar\theta$ of the supermanifold
$$
\begin{array}{lcl} 
(B_\mu^a T^a) (x, \bar\theta) &=& (A_\mu^a T^a) (x) + \bar \theta \;
(R_\mu^a T^a) (x)\nonumber\\
(\Phi^a T^a) (x, \bar\theta)  &=& (C^a T^a) (x) 
- i \; \bar\theta\; ({\cal B}^a T^a) (x) \nonumber\\
(\bar \Phi^a T^a) (x, \bar\theta) &=& (\bar C^aT^a) (x)  
+ i\;\bar\theta\; (B^aT^a) (x)
\end{array} \eqno(3.1)
$$
where the signs in the above expansions have been chosen for the later 
algebraic convenience only. Some of the salient and relevant points at this 
juncture are: (i) in general, the 
$(2+2)$-dimensional supermanifold is parametrized by the superspace coordinates
$Z^M = (x^\mu,\theta,\bar\theta)$ where $x^\mu (\mu = 0, 1)$ are the even
spacetime coordinates and $\theta, \bar \theta$ are the odd Grassmannian
variables ($\theta^2 = \bar \theta^2 = 0, \theta\bar\theta 
+ \bar\theta\theta = 0$). However, here we choose only the chiral
superfields which depend only on 
the superspace variables $Z^M = (x^\mu, \bar\theta)$. (ii) The most general
expansions for the even superfield $B_\mu$ and odd superfields $\Phi$
and $\bar \Phi$, along $\theta$, $\bar\theta$ and $\theta\bar\theta$ 
directions of the supermanifold,  are [17,21]
\footnote{ All the signs in the analogue of the super
expansions (3.2) have been taken to be 
{\it only} positive in Refs. [17,21]. We invoke here some negative signs for 
the sake of later algebraic convenience without changing the 
physical contents of the theory.}
$$
\begin{array}{lcl} 
(B_\mu^a T^a) (x, \bar\theta) &=& (A_\mu^a T^a) (x) + \bar \theta \;
(R_\mu^a T^a) (x) + \theta\; (\bar R_\mu^a T^a) (x)
+ i \;\theta\bar\theta (S_\mu^aT^a) (x)\nonumber\\
(\Phi^a T^a) (x, \bar\theta)  &=& (C^a T^a) (x) 
- i \; \bar\theta\; ({\cal B}^a T^a) (x) + i\;\theta\; (\bar B^aT^a) (x)
+ i\; \theta\bar\theta (s^aT^a) (x) \nonumber\\
(\bar \Phi^a T^a) (x, \bar\theta) &=& (\bar C^aT^a) (x)  
+ i\;\bar\theta\; (B^aT^a) (x) - i\;\theta\; (\bar {\cal B}^aT^a) (x)
+ i \; \theta\bar\theta\; (\bar s^aT^a)(x).
\end{array} \eqno(3.2)
$$
The chiral limit (i.e. $\theta \rightarrow 0$) of the above expansion
have been taken into (3.1). (iii) The horizontality condition 
(see, e.g., Refs. [17,21])  on
(3.2) leads to the derivation of the off-shell nilpotent (anti-)BRST
symmetries. We shall see later that the same condition on (3.1) yields the
on-shell nilpotent BRST symmetry {\it alone}. (iv) The auxiliary
fields in (3.1) are the fermionic (odd) fields $R_\mu$ and 
bosonic (even) fields
$B$ and ${\cal B}$. The corresponding fields in (3.2) are:
$R_\mu, \bar R_\mu, s, \bar s$ and $B, \bar B, {\cal B}, \bar {\cal B},
S_\mu$. (v) In the above expansions (3.1) and (3.2), all the local fields
on the r.h.s. are functions of the spacetime variables $x^\mu$ alone.

Now we invoke the horizontality condition ($\tilde F = \tilde D \tilde A
= D A =  F$) on the super curvature two-form
$\tilde F = \frac{1}{2}\;(dZ^M \wedge dZ^N) \tilde F_{MN}$ by
exploiting the Maurer-Cartan equation, as 
$$
\begin{array}{lcl} 
\tilde F = \tilde d \tilde A + \tilde A \wedge \tilde A \equiv 
d A + A \wedge A = F
\end{array} \eqno(3.3)
$$
where the super one-form connection $\tilde A$ (in terms of the {\it chiral}
superfields) 
and super exterior derivative
$\tilde d$ (in terms of the {\it chiral} superspace variables 
$(x^\mu, \bar\theta)$), are defined as
$$
\begin{array}{lcl}
\tilde A &=& d Z^M\; \tilde A_{M} = d x^\mu \;B_{\mu} (x , \bar \theta)
+ d \bar \theta\; \Phi ( x, \bar \theta) \nonumber\\
\tilde d &=& \;d Z^M \;\partial_{M} = d x^\mu\; \partial_\mu\;
+ \;d \bar \theta \;\partial_{\bar \theta}.
\end{array}\eqno(3.4)
$$
The above definitions lead to the following expressions for
$$
\begin{array}{lcl} 
\tilde d \tilde A &=& (dx^\mu \wedge dx^\nu) (\partial_\mu B_\nu)
+ (dx^\mu \wedge d \bar\theta) (\partial_\mu \Phi 
- \partial_{\bar\theta} B_\mu) - (d \bar\theta \wedge d \bar\theta)
(\partial_{\bar\theta} \Phi) \nonumber\\
\tilde A \wedge \tilde A &=&
(dx^\mu \wedge dx^\nu) (B_\mu B_\nu)
+ (dx^\mu \wedge d \bar\theta) ( [ B_\mu,  \Phi] ) 
- \frac{1}{2}\;(d \bar\theta \wedge d \bar\theta) (\{ \Phi, \Phi \}).
\end{array} \eqno(3.5)
$$
Now the horizontality restrictions in (3.3) imply the following
$$
\begin{array}{lcl}
&&\partial_\mu \Phi - \partial_{\bar\theta} B_\mu + [ B_\mu, \Phi ] = 0,
\quad \partial_{\bar\theta} \Phi + \frac{1}{2}\; \{ \Phi, \Phi \} = 0
\nonumber\\
&&\partial_\mu B_\nu - \partial_\nu B_\mu + [ B_\mu, B_{\nu} ] =
\partial_\mu A_\nu - \partial_\nu A_\mu + [ A_\mu, A_{\nu} ].
\end{array} \eqno(3.6)
$$
The first two relations in the above equation lead to the following 
expressions for the auxiliary fields in terms of the basic fields of the
Lagrangian density (2.1)
$$
\begin{array}{lcl}
R_{\mu} \;(x) &=& D_{\mu}\; C(x)\; \quad 
{\cal B} (x) = -  \frac{i}{2}\; (C \times C)(x) 
\qquad [\; {\cal B}(x),  C(x)\; ] = 0
\end{array} \eqno(3.7)
$$
and the l.h.s. of
the last relationship (with $\tilde F_{\mu\nu} = \partial_\mu B_\nu
- \partial_\nu B_\mu + [B_\mu, B_\nu ] $) yields
$$
\begin{array}{lcl}
\tilde F_{\mu\nu} = F_{\mu\nu} + \bar\theta\; (D_\mu R_\nu - D_\nu R_\mu )
\equiv F_{\mu\nu} + \bar\theta\; F_{\mu\nu} \times C 
\end{array} \eqno(3.8)
$$
where we have used $D_\mu R_\nu = \partial_\mu R_\nu + [A_\mu, R_\nu],
\;R_\mu = D_\mu C,\; [ D_\mu, D_\nu ]\; C = F_{\mu\nu} \times C$. The total 
antisymmetric property of $f^{abc}$ in $a, b, c$ allows one to trivially
check that the kinetic energy term of the Lagrangian density (2.1)
remains invariant under transformation (3.8) (i.e. $- \frac{1}{4}
F^{\mu\nu} \cdot F_{\mu\nu} = - \frac{1}{4}\; \tilde F^{\mu\nu}
\cdot \tilde F_{\mu\nu}) $. Thus, physically, the horizontality condition 
implies that the kinetic energy term of the Lagrangian density remains
invariant (and equal to the square of the {\it ordinary} curvature tensor).
In other words, the supersymmetric contribution coming from the
$\bar\theta$-component of the super curvature tensor $\tilde F_{\mu\nu}$
{\it does not} lead to any additional changes to the usual
kinetic energy term ($- \frac{1}{4} F^{\mu\nu} \cdot F_{\mu\nu}$) which
is defined on the ordinary 2D spacetime manifold. It is obvious from equation 
(3.7) that the horizontality restriction (3.3) does not fix the auxiliary 
field $B (x) = (B^a T^a) (x)$ in terms of the basic fields of the 
Lagrangian density (2.1). However,
it has been shown [5,6] that the off-shell nilpotent BRST- and co-BRST 
symmetries can  be derived if we linearize the kinetic- and gauge-fixing
terms of (2.1) by invoking two auxiliary fields
$ {\cal B}$ and $B$ in the following way
$$
\begin{array}{lcl}
{\cal L}_{{\cal B}} = {\cal B} \cdot E - \frac{1}{2} {\cal B} \cdot {\cal B}
+ (\partial_\rho A^\rho ) \cdot B + \frac{1}{2} B \cdot B - i \partial_\mu
\bar C \cdot D^\mu C
\end{array} \eqno(3.9)
$$
which shows that $ B = - (\partial_\rho A^\rho) $
\footnote{ It is clarifying to note that the 
off-shell nilpotent BRST ($\tilde s_{b}$)- and co-BRST ($\tilde s_{d}$)
symmetry transformations: $\tilde s_{b} A_\mu = D_\mu C, \;
\tilde s_{b} \bar C = i B, \;\tilde s_{b} B = 0,\;
\tilde s_{b} C = - \frac{1}{2} C \times C, 
\;\tilde s_{b} {\cal B} = {\cal B} \times C, \;\;\tilde s_{b} E 
= E \times C, \; \tilde s_{b} (\partial_\rho A^\rho) =
\partial_\rho D^\rho C$ and  $ \tilde s_{d} A_\mu 
= - \varepsilon_{\mu\nu} \partial^\nu \bar C,\; 
\tilde s_{d} C = - i {\cal B}, \;\tilde s_{d} B = 0,\; 
\tilde s_{d} \bar C = 0,\; 
\tilde s_{d} {\cal B} = 0, \;\tilde s_{d} (\partial_\rho A^\rho) = 0, 
\;\tilde s_{d} E = D_\mu \partial^\mu \bar C,\;\;
\tilde s_{d} (D_\mu \partial^\mu \bar C) = 0$ leave
the Lagrangian density (3.9) invariant
(up to a total derivative).}. Thus, the super expansion
in (3.1) can be re-expressed, in terms of 
the expressions for the auxiliary fields in (3.7) and
$B = - (\partial_\rho A^\rho)$, as
$$
\begin{array}{lcl} 
B_\mu (x, \bar\theta) &=& A_\mu (x) + \bar \theta \;
(D_\mu C) (x) \equiv A_\mu (x) + \bar\theta\; (s_{b} A_\mu (x))
\nonumber\\
\Phi (x, \bar\theta)  &=& C (x) 
- \frac{1}{2}\; \bar\theta\; (C \times C) (x) \equiv C (x) + \bar\theta\; 
(s_{b} C (x)) \nonumber\\
\bar \Phi (x, \bar\theta) &=& \bar C (x)  
- i\;\bar\theta\; (\partial_\rho A^\rho) (x) \equiv \bar C (x) + \bar\theta\;
(s_{b} \bar C (x)).
\end{array} \eqno(3.10)
$$
With the above expansions as inputs, the on-shell nilpotent BRST symmetries
in (2.2) can be concisely expressed in the language of superfields as
$$
\begin{array}{lcl}
s_{b} \;B_{\mu} 
= \partial_{\mu} \Phi  + (B_\mu \times \Phi)\qquad
\;s_{b}\; \Phi  = -\; \frac{1}{2} (\Phi \times \Phi)\qquad
\; s_{b}\; \bar \Phi  = -i\; (\partial_\mu B^\mu).
\end{array}\eqno(3.11)
$$
In fact, in the above {\it three} transformations, the first one yields
$s_{b} A_\mu = D_\mu C,\; s_{b} C = - \frac{1}{2} C \times C$; the second
produces $ s_{b} C = - \frac{1}{2} C \times C $ and the third leads to
$s_{b} \bar C = - i (\partial_\mu A^\mu), \;
s_{b} (\partial_\mu A^\mu) = \partial_\mu D^\mu C$ in terms of the basic
fields of (2.1). Comparing with (2.4), it is clear that
$$
\begin{array}{lcl}
i\; {\displaystyle \frac{\partial}{\partial \bar\theta}}\; B_\mu (x,\bar\theta)
= [ Q_{b}, A_\mu ] \qquad
i\;{\displaystyle \frac{\partial}{\partial \bar\theta}}\; \Phi (x,\bar\theta)
= \{ Q_{b}, C \} \qquad
i\;{\displaystyle 
\frac{\partial}{\partial \bar\theta}}\; \bar \Phi (x,\bar\theta)
= \{ Q_{b}, \bar C \}
\end{array}\eqno(3.12)
$$
which shows that the conserved and on-shell  nilpotent BRST charge $Q_{b}$
(that generates the BRST transformations (2.2)) can be 
{\it geometrically } interpreted as
the generator of translation ($\frac{\partial}{\partial\bar\theta}$)
along the Grassmannian direction $\bar\theta$ of the supermanifold. This
clearly establishes the fact that the horizontality condition w.r.t.
the super covariant derivative $\tilde D = \tilde d + \tilde A$
(in $ \tilde F = \tilde D \tilde A = D A = F$) leads to the derivation
of the on-shell nilpotent BRST symmetries for the non-Abelian gauge theory
and it plays an important role in providing the geometrical interpretation
for the BRST charge $Q_{b}$ on the $(2+2)$-dimensional supermanifold.\\

\noindent
{\bf 4 On-shell nilpotent co-BRST symmetry: superfield approach}\\

\noindent
It is evident
 from our earlier discussions that $F = D A = d A + A \wedge A$
defines the two-form curvature tensor on an ordinary compact manifold. The
operation of an ordinary co-exterior derivative $\delta = - * d *$ on the
one-form $A = dx^\mu A_\mu^a T^a$ leads to the definition of the gauge-fixing
term (i.e. $\delta A = \partial_\mu A^{\mu a} T^a$). Interestingly, the
operation of the covariant co-exterior derivative $\Omega = - * D * $
on the one-form $A$ leads to the same gauge-fixing term. This can be
seen (with $ * (d x^\mu) = \varepsilon^{\mu\nu} (dx_\nu),  
* (dx^\mu \wedge dx^\nu) = \varepsilon^{\mu\nu}, * A = \varepsilon^{\mu\nu}
(dx_\nu) A_\mu$) as follows
$$
\begin{array}{lcl}
\Omega A = - * (d + A) * A = - * d (* A)  - * (A \wedge * A) =
\partial_\mu A^{\mu a} T^a 
\end{array} \eqno(4.1)
$$
where the total
antisymmetry property of the $f^{abc}$ plays an important role in
proving $A_\mu \times A^\mu = 0$. In the simplest way, this statement can be
verified by noting that: $D_\mu A^\mu = \partial_\mu A^\mu 
+ A_\mu \times A^\mu = \partial_\mu A^\mu$ which is, in essence, the
reflection of (4.1).

Now we shall generalize the horizontality condition ($\tilde D \tilde A = D A$)
of equation (3.3) (where (super) exterior derivatives $(\tilde d) d$ and
(super) one-forms $(\tilde A) A$ play an important role) to the case where
the (super) co-exterior derivatives $(\tilde \delta) \delta$ operate on
(super) one-forms $(\tilde A) A$
to define a (super-)scalar. Thus, the analogue of
the horizontality condition 
\footnote{ This condition has been christened as the dual horizontality
condition in Refs. [18,26] because the (super) co-exterior derivatives
$(\tilde \delta) \delta$ are Hodge dual to (super) exterior derivatives
$(\tilde d) d$ on the (super) manifolds with (super) Hodge operations
$(\star) *$ as defined in (4.2) and (4.4).} is 
$$
\begin{array}{lcl} 
\tilde \delta \tilde A = \delta A \quad
\delta A = (\partial_\mu A^\mu) \quad
\delta = - * d * \quad  \tilde \delta = - \star \tilde d \star
\quad A  = dx^\mu A_\mu 
\end{array} \eqno(4.2)
$$
where, in the definition of $\tilde \delta \tilde A 
= - \star \tilde d \star \tilde A$, we have to take into account
$$
\begin{array}{lcl}
\tilde d =  d x^\mu\; \partial_\mu\;
+ \;d \bar \theta \;\partial_{\bar \theta} \qquad
\star\; \tilde A = \varepsilon^{\mu\nu} (d x_\nu) \;B_{\mu} (x , \bar \theta)
+ d \bar \theta\; \bar \Phi ( x, \bar \theta) 
\end{array}\eqno(4.3)
$$
so that the operation of $\tilde d$ on the one-form $(\star \tilde A)$
can exist in the chiral space. Here the Hodge duality $\star$ operation
is defined on the $(2+2)$-dimensional supermanifold. In its most general form,
this operation on the super differentials and their wedge products, are
$$
\begin{array}{lcl}
\star\; (dx^\mu) &=& \varepsilon^{\mu\nu}\; (d x_\nu)\; \qquad \;
\star\; (d\theta) = (d \bar \theta)\; \qquad \;
\star\; (d \bar \theta) = (d \theta) \nonumber\\
\star\; (d x^\mu \wedge d x^\nu) &=& \varepsilon^{\mu\nu}\; \qquad 
\star\; (dx^\mu \wedge d \theta) = \varepsilon^{\mu\theta}\; \qquad
\star\; (dx^\mu \wedge d \bar \theta) = \varepsilon^{\mu\bar\theta}
\nonumber\\
\star \; (d \theta \wedge d \theta) &=& s^{\theta\theta} \; \qquad
\star \; (d \theta \wedge d \bar \theta) = s^{\theta \bar \theta} \; \qquad
\star \; (d \bar \theta \wedge d \bar \theta) = s^{\bar \theta \bar \theta}
\end{array} \eqno(4.4)
$$
where $\varepsilon^{\mu\theta} = - \varepsilon^{\theta\mu}, \varepsilon^{\mu
\bar\theta} = - \varepsilon^{\bar\theta \mu},  
s^{\theta \bar \theta} = s^{\bar\theta\theta}$ etc. 
The choice of $(\star \tilde A$) in (4.3) is derived from the following
general expression for the super one-form $\tilde A$ and the application of 
the $\star$ operation (4.4) on it, in the $(2+2)$-dimensional supermanifold:
$$
\begin{array}{lcl}
\tilde A (x,\theta, \bar\theta) &=&
 dx^\mu B_\mu (x,\theta,\bar\theta) 
+ d \theta \;\bar \Phi (x,\theta,\bar\theta)
+ d \bar \theta \; \Phi (x,\theta,\bar\theta) \nonumber\\
\star\; \tilde A (x,\theta,\bar\theta)
&=& \varepsilon^{\mu\nu} (dx_\nu) B_\mu (x,\theta,\bar\theta)
+ d \bar \theta \;\bar \Phi (x,\theta,\bar\theta)
+ d \theta \; \Phi (x,\theta,\bar\theta).
\end{array} \eqno(4.5)
$$
Taking the chiral limit ($\theta \rightarrow 0, d \theta \rightarrow 0$)
of the above equation leads to the 
proof for the choice of $(\star \tilde A)$ in (4.3).
With the help of (4.4) and (4.3), the l.h.s. of the dual
horizontality condition (4.2), can be explicitly written as
$$
\begin{array}{lcl}
\tilde \delta \tilde A = (\partial_\mu B^\mu) +  
s^{\bar \theta \bar\theta} (\partial_{\bar\theta}
\bar \Phi) 
- \varepsilon^{\mu \bar \theta} (\partial_{\mu}
\bar \Phi + \varepsilon_{\mu\nu} \partial_{\bar \theta} B^\nu).
\end{array} \eqno(4.6)
$$
Application of the requirement in (4.2) allows us to set the
coefficients of $\varepsilon^{\mu\bar\theta}$ and $s^{\bar\theta\bar\theta}$
equal to zero. This restriction leads to the following relationships
$$
\begin{array}{lcl}
R_{\mu} (x) = - \varepsilon_{\mu\nu} \partial^\nu \bar C (x)
\qquad B (x) = 0.
\end{array} \eqno(4.7)
$$
The other restriction, ensuing from (4.2), is $\partial_\mu B^\mu =
\partial_\mu A^\mu$ which leads to $\partial_\mu R^\mu = 0$. It is evident
that (4.7) automatically satisfies this condition. Physically, the dual 
horizontality condition amounts to the restriction that the ordinary
gauge-fixing term $(\partial \cdot A)$ defined on the ordinary 2D spacetime
manifold remains intact and unchanged. In other words, the supersymmetric
contribution emerging from the coefficients $\varepsilon^{\mu\bar\theta},
s^{\bar\theta\bar\theta}$ in (4.6) do not alter the original 
value of the gauge-fixing term defined on the 2D ordinary
spacetime manifold (i.e. $\delta A = (\partial\cdot A)$).
It will be noticed that the
auxiliary field ${\cal B} (x)$ is not fixed by the dual
horizontality condition in (4.2). However, our argument in the context
of choice of the Lagrangian density (3.9) for the off-shell nilpotent BRST-
and co-BRST symmetries, comes to our rescue as we have:
$ {\cal B} (x) = E (x)$. Thus, the expansion (3.1), 
with the results in (4.7), can be expressed as
$$
\begin{array}{lcl}
B_{\mu}\; (x, \bar \theta) &=& A_{\mu} (x) 
- \;\bar \theta\;\varepsilon_{\mu\nu}\; \partial^{\nu} \bar C (x) 
\equiv A_\mu (x)  + \; \bar \theta\;
(s_{d} A_\mu (x)) \nonumber\\
\Phi\; (x, \theta, \bar \theta) &=& C (x) 
- \;i \;\bar \theta \; E (x)\; \equiv\; C (x) + \;\bar \theta\; (s_{d} C(x))
\nonumber\\
\bar \Phi\; (x, \theta, \bar \theta) &=& \bar C (x) 
+\; i \; \bar \theta\; (B (x) = 0) 
\equiv \bar C (x) + \;\bar\theta\; (s_{d} \bar C(x)).
\end{array} \eqno(4.8)
$$
The above expansions (due to the dual horizontality condition
in (4.2)) allow one to express the on-shell nilpotent co-BRST symmetry
transformations of (2.2), in terms of the chiral superfields (4.8), as
$$
\begin{array}{lcl}
s_{d}\; B_{\mu} 
= - \varepsilon_{\mu\nu} \partial^{\nu} \bar \Phi\; \qquad
\;s_{d}\; \bar \Phi  = 0\;\qquad
 s_{d} \; \Phi 
= + i\; \varepsilon^{\mu\nu} (\partial_\mu 
B_\nu + \frac{1}{2}\; B_\mu \times B_\nu)
\end{array}\eqno(4.9)
$$
where the first transformation in the above equation leads to
$s_{d} A_\mu = - \varepsilon_{\mu\nu} \partial^\nu \bar C, s_{d} \bar C = 0$;
the second one yields $s_{d} \bar C = 0$ and third one produces
$s_{d} C = - i E,\; s_{d} E = D_\mu \partial^\mu \bar C$ in the language
of the basic fields of the Lagrangian density (2.1).
The equation (4.8)
establishes that the on-shell nilpotent co-BRST charge $Q_{d}$
geometrically corresponds to the translation generator 
$(\frac{\partial} {\partial \bar\theta})$ along the Grassmannian direction
$\bar\theta$ of the supermanifold as
$$
\begin{array}{lcl} 
{\displaystyle \frac{\partial}{\partial\bar\theta}} \Sigma (x,\bar\theta)
= - i\; [ \Lambda (x), Q_{d} ]_{\pm} \qquad \Sigma = \Phi, \bar \Phi, B_\mu
\quad \Lambda = C, \bar C, A_\mu
\end{array} \eqno(4.10)
$$
where the bracket $[\;,\;]_{\pm}$ stands for the (anti-)commutator for
the $\Sigma$ (or corresponding $\Lambda$) being (fermionic)bosonic in nature
and we have exploited the defining relationship (2.4).\\

\noindent
{\bf 5 Topological aspects: superfield formalism}\\

\noindent
We have derived in section 2 some of the key topological features in the 
Lagrangian formulation and have shown that, modulo some total derivatives,
the Lagrangian density (2.1) can be expressed as the sum of BRST- and co-BRST
anti-commutator (cf. (2.5)). In the language of the chiral superfields,
the same can be expressed, modulo total derivative $\partial_\mu X^\mu$, as:
$$
\begin{array}{lcl} 
{\cal L}_{B} = - {\displaystyle
\frac{i}{2}\;\frac{\partial}{\partial \bar \theta}}\;
\Bigl ( \; [ \varepsilon^{\mu\nu} (\partial_{\mu} B_\nu + \frac{1}{2}
B_\mu \times B_\nu) \cdot
\Phi ]|_{\mbox{co-BRST}}  +  [
(\partial_\mu B^\mu) \cdot \bar \Phi ]|_{\mbox{BRST}}\;  \Bigr )
\end{array} \eqno(5.1)
$$
where the subscripts BRST- and co-BRST stand for the insertion
of the chiral super expansions given in (3.10) and (4.8) respectively
and $X^\mu = \frac{i}{2} (\bar C \cdot D^\mu C + \partial^\mu \bar C
\cdot C)$. In the above computation,  we have used
$$
[ \varepsilon^{\mu\nu} (\partial_{\mu} B_\nu  + \frac{1}{2}
B_\mu \times B_\nu) \cdot \Phi ]|_{\mbox{co-BRST}} = - (E \cdot C
+ \bar\theta\; D_\mu \partial^\mu \bar C \cdot C) + i\; \bar\theta\;
E \cdot E.
$$
Mathematically, the above Lagrangian density is nothing but the 
$\bar\theta$-component of 
the composite fields $(\partial_\mu B^\mu)\cdot \bar \Phi$
and $\varepsilon^{\mu\nu} (\partial_{\mu} B_\nu + \frac{1}{2}
B_\mu \times B_\nu) \cdot\Phi $ when we substitute the chiral expansions
(3.10) and (4.8) that have been obtained after the application of 
(dual) horizontality conditions. Incorporating the geometrical
interpretation for the on-shell nilpotent (co-)BRST charges, it can seen
that the Lagrangian density (2.1) corresponds to the translation of some
local (but composite) chiral superfields along the $\bar\theta$-direction
of the $(2+2)$-dimensional supermanifold where the generators of translation
on the suprmanifold are conserved and on-shell nilpotent
BRST- and co-BRST charges ($Q_{b}$ and $Q_{d}$).

Let us now concentrate on the topological invariants of the theory. We can
provide the geometrical origin for the zero-forms $W_{0}$ and $V_{0}$
of equations (2.10) and (2.9) which are (co-)BRST invariants. To this end in
mind, we note the following
$$
\begin{array}{lcl}
(\Phi \cdot \bar \Phi)|_{\mbox{BRST}}
&=& C \cdot \bar C + i\; \bar \theta\;
C \cdot (\partial_\mu A^\mu) - \frac{1}{2}\;
\bar\theta\; (C \times C) \cdot \bar C \nonumber\\
(\Phi \cdot \bar \Phi)|_{\mbox{co-BRST}}
&=& C \cdot \bar C - i \;\bar \theta\;
E \cdot \bar C.
\end{array} \eqno(5.2)
$$
It obvious now that zero-forms in (2.9) and (2.10) are as follows
$$
\begin{array}{lcl}
i\; {\displaystyle \frac{\partial} 
{\partial \bar \theta}}\;
(\Phi\cdot \bar\Phi)|_{\mbox{BRST}} = V_{0} \qquad
i\; {\displaystyle \frac{\partial} 
{\partial \bar \theta}} 
(\Phi\cdot \bar\Phi)|_{\mbox{co-BRST}}
= W_{0}.
\end{array} \eqno(5.3)
$$
Mathematically, it means that the on-shell ($\partial_\mu D^\mu C = 0$)
BRST invariant quantity $V_{0}$ is nothing but the $\bar\theta$-component
of the local (but composite) chiral superfield $(\Phi \cdot \bar \Phi)$
when we substitute for them the super expansions
(3.10) that are obtained after the imposition of
the horizontality condition (3.3). In the language of the geometry on the
supermanifold, $V_{0}$ is equivalent to a translation 
of the chiral superfield $(\Phi \cdot \bar\Phi)$ along 
the $\bar\theta$-direction 
which is generated by the on-shell nilpotent  BRST charge $Q_{b}$.
In a similar fashion, we can provide a geometrical interpretation to the
on-shell ($D_\mu\partial^\mu \bar C = 0$) co-BRST invariant zero-form $W_{0}$.
The rest of the topological invariants $(V_{k}, W_{k}, k = 1, 2)$ can
be computed by the following recursion relations [5,6] that characterize 
the topological nature of this theory:
$$
\begin{array}{lcl}
s_{b}\; V_{k} = d\; V_{k-1} \qquad
s_{d}\; W_{k} = \delta\; W_{k-1} \qquad
d = dx^\mu \partial_\mu, \qquad
\delta = i\;dx^\mu \varepsilon_{\mu\nu} \partial^{\nu}. 
\end{array} \eqno(5.4)
$$

Now we wish to provide the geometrical interpretation for the symmetric
energy-momentum tensor $T_{\mu\nu}^{(s)}$
of the theory in the language of the translation
on the four $(2+2)$-dimensional supermanifold. In fact, it can be checked
that the $T_{\mu\nu}^{(s)}$ of (2.6), modulo some total derivatives
$X_{\mu\nu}^{(s)}$, can be expressed as
$$
\begin{array}{lcl}
&&T_{\mu\nu}^{(s)} + X_{\mu\nu}^{(s)}= {\displaystyle \frac{i}{2}\;
\frac{\partial} {\partial \bar \theta}} \; \Bigl (\; 
[\; Y_{\mu\nu}^{(s)}\;]|_{\mbox{BRST}}
+ [\;Z_{\mu\nu}^{(s)}\;]|_{\mbox{co-BRST}}\; \Bigr ) \nonumber\\
&& Y_{\mu\nu}^{(s)} = 
\partial_\mu \bar \Phi \cdot B_\nu
+ \partial_\nu \bar \Phi \cdot B_\mu + \eta_{\mu\nu} (\partial_\rho A^\rho)
\cdot \bar \Phi  \nonumber\\
&& Z_{\mu\nu}^{(s)} = 
\varepsilon_{\mu\rho} \partial_\nu \Phi \cdot B^\rho
+ \varepsilon_{\nu\rho} \partial_\mu \Phi \cdot B^\rho + \eta_{\mu\nu}
\varepsilon^{\rho\sigma} (\partial_\rho B_\sigma + \frac{1}{2} \;B_{\rho}
\times B_\sigma) \cdot \Phi
\end{array}\eqno(5.5)
$$
where the explicit form of the total derivative term $X_{\mu\nu}^{(s)}$ is
$$
\begin{array}{lcl}
X_{\mu\nu}^{(s)} &=& 
\frac{1}{2} \; \partial_\mu [ (\partial_\rho A^\rho)
\cdot A_\nu + E \cdot \varepsilon_{\nu\rho} A^\rho ]
+ \frac{1}{2} \; \partial_\nu [ (\partial_\rho A^\rho)
\cdot A_\mu + E \cdot \varepsilon_{\mu\rho} A^\rho ] \nonumber\\
&-& \frac{i}{2}\; \eta_{\mu\nu}\; \partial_\rho [\;\partial^\rho \bar C \cdot C
+ \bar C \cdot D^\rho C\;].
\end{array} \eqno(5.6)
$$
It is obvious from (5.5) that $T_{\mu\nu}^{(s)}$ geometrically corresponds to
the translations of the local (but composite) chiral superfields 
$Y_{\mu\nu}^{(s)}$
and $Z_{\mu\nu}^{(s)}$ along the $\bar\theta$-direction of the four
$(2+2)$-dimensional supermanifold. These translations are generated by the
on-shell nilpotent BRST- and co-BRST charges $Q_{b}$ and $Q_{d}$ 
respectively. Mathematically, the expression for $T_{\mu\nu}^{(s)}$,
modulo some total derivatives, is nothing but the $\bar\theta$-component
of the composite chiral superfields $Y_{\mu\nu}^{(s)}$ and $Z_{\mu\nu}^{(s)}$
where the expansions (4.8) and (3.10) for the chiral superfields 
are taken into account. Of course, these expansions are obtained
after the imposition of (dual) horizontality conditions which play
a very significant role here.\\

\noindent
{\bf 6 Conclusions}\\

\noindent
In our present investigation, we have concentrated on 
the key topological properties of the 2D self-interacting non-Abelian
gauge theory in the framework of the {\it chiral} superfield formulation.
Our key observations are: (i) it is the existence of the
novel on-shell nilpotent co-BRST symmetry (together with 
the familiar on-shell nilpotent BRST symmetry) that enables us to furnish 
a convincing proof for the topological nature of the 2D self-interacting 
non-Abelian gauge theory in the Lagrangian formulation because the
Lagrangian density and symmetric energy momentum tensor for the
theory turn out to be the sum of BRST- and co-BRST invariant
parts (cf. (2.5) and (2.7)). 
(ii) In the framework of superfield 
formulation, this fact is reflected in the appearance of the Lagrangian
density and the symmetric energy momentum tensor which turn out to be the
total derivative w.r.t. Grassmannian variable $\bar\theta$
(cf. (5.1) and (5.5)). Geometrically,
this is equivalent to the translation of some local (but composite) chiral 
superfields along the $\bar\theta$-direction of the supermanifold. These
translations are basically generated by the conserved and on-shell nilpotent 
(co-)BRST charges. (iii) Our claim is that
whenever a Lagrangian density and corresponding symmetric energy-momentum
tensor turn out to be a total derivative w.r.t. Grassmannian variable,
the theory is a topological field theory and it owes its origin to the
(super) cohomological operators $(\tilde d) d $ and/or
 $(\tilde \delta) \delta$. (iv) It is important for our whole discussion 
(in our present paper) to derive the on-shell nilpotent
(co-)BRST symmetries in the superfield formulation 
because the Lagrangian density (2.1) is endowed with
{\it only} these symmetries and it does not respect
anti-BRST- and anti-co-BRST symmetries. 
(v) The choice of the {\it chiral}
superfields and imposition of the (dual) horizontality
conditions enable us to derive the on-shell nilpotent BRST- and co-BRST 
symmetries. This feature of our present investigation is different from the 
earlier attempts to derive the off-shell nilpotent (anti-)BRST symmetries 
[20-22,17] (and (anti-)co-BRST symmetries [17]) in the 
framework of superfield formulation where
the most general super expansion for the superfields 
were considered. (vi) In our present analyses, 
the Lagrangian density (2.1) and corresponding symmetric energy momentum 
tensor (2.6) play  key roles. Thus, the geometrical understanding of
these physically relevant and interesting quantities {\it might} turn out to
play an important role in the context of 2D gravity where a non-trivial
(spacetime dependent) metric is chosen for the discussion of such 
gauge theories in the background of the curved spacetime (super)manifolds.

It is a well-known fact that the 2D Abelian as well as non-Abelian
(Yang-Mills) gauge theories possess no physical degrees of
freedom when they are defined on an ordinary 2D spacetime
manifold without any non-trivial topology at the boundary. In other words, 
for all such manifolds, the
fields of the theory are assumed to fall-off rapidly at infinity. Thus, our
present 2D gauge theory, according to the
standard definition of a TFT on a flat spacetime manifold with a spacetime
independent metric (see, e.g. [1] for details), turns out to be a new type
of TFT because the (co-)BRST symmetries of the theory gauge out the propagating
degrees of freedom. In our chiral superfield formulation, this
fact is reflected in (5.1) and (5.5) where we have been able to show 
that the Lagrangian density and symmetric energy momentum tensor of the theory
are total Grassmannian derivatives modulo some total spacetime derivatives.
These latter derivatives do not contribute anything substantial in our 
present discussion. This will not be the case, however, if this 2D 
gauge theory is defined, say, on a circle where the boundary terms will 
contribute. In fact for this case, there will
be physical degrees of freedom associated with the gauge field. Furthermore,
it is clear that all the physical fields of the theory 
will not go to zero at the boundary. Consequently,
this theory will {\it not} be a TFT. In mathematical terms, now
the total derivatives $\partial_\mu X^\mu$ and $X^{(s)}_{\mu\nu}$ of (5.1) and
(5.5) cannot be neglected. As a result, the Lagrangian
density and symmetric energy momentum tensor will {\it not}
be able to be expressed
as the sum of (co-) BRST anti-commutators (in contrast to what we have   
shown in (2.5) and (2.7)). In the language of our present chiral superfield 
formulation, we shall {\it not} be able to express the above quantities
as a total derivative w.r.t. the Grassmannian variable.

It would be nice to generalize our present work to 4D two-form free
Abelian gauge theory where the existence of (co-)BRST symmetries has
been shown [10]. On its face value, however, it appears that there will
be difficulty in such a generalization because of the fact that there is
a single physical degree of freedom associated with the gauge field of the
theory. But, we feel, it is important to try such a generalization so that
we can learn more about some aspects of
the two-form gauge theory which plays such
a significant role in the context of string theory
as well as field theory. In fact, some steps have already been taken 
in this direction [33]. The superfield  formulation of the 2D interacting 
(non-)Abelian gauge theories is another direction that can be 
pursued later. It might be interesting to follow the approach adopted
in [32] to discuss the 2D free Abelian- and self-interacting non-Abelian
gauge theories on the 2D closed Riemann surface (i.e. the Euclidean version of 
the 2D Minkowski manifold) of genus-one (and/or higher genus
Riemann surfaces) and study the topological 
invariants of this theory. It would be gratifying to find their
connection with the pertinent notions
in the domain of algebraic geometry. These
are some of the issues that are under investigation and our results will
be reported elsewhere.\\

\noindent
{\bf Acknowledgements:}
Clarifying comments by the referees are gratefully acknowledged.

\end{document}